\documentclass[nofootinbib,floatfix,prd,preprintnumbers,superscriptaddress]{revtex4}
\usepackage{graphicx}
\usepackage{amsfonts}
\usepackage{amssymb}
\usepackage{amsbsy}
\usepackage{amsmath}
\usepackage{latexsym}
\usepackage{bm}
\usepackage{hyperref}

\newcommand*{\di}{\partial}

\newcommand*{\rhohat}{\hat{\rho}}

\renewcommand*{\c}{\text{c}}

\newcommand*{\GR}{\text{\tiny{GR}}}
\newcommand*{\KK}{\mathcal{E}}
\newcommand*{\Weyl}{\mathcal{E}}

\newcommand*{\etahat}{\hat{\eta}}

\renewcommand*{\a}{\hat{a}}
\renewcommand*{\b}{\text{b}}
\renewcommand*{\k}{\hat{k}}

\renewcommand*{\H}{\hat{H}}

\newcommand*{\approaches}[2]{\xrightarrow[#2]{\,\,\,{#1}\,\,\,}}
\newcommand*{\PS}{\text{\tiny{PS}}}
\newcommand*{\CI}{\text{\tiny{CI}}}
\newcommand*{\eff}{\text{\tiny{eff}}}
\newcommand*{\fiveD}{\text{\tiny{5D}}}
\newcommand*{\prim}{\text{\tiny{inf}}}

\begin{document}

\preprint{arXiv:0705.1685 [astro-ph]}

\title{Scalar perturbations in braneworld cosmology}

\author{Antonio Cardoso}%
\email{antonio.cardoso.AT.port.ac.uk}%
\affiliation{Institute of Cosmology \& Gravitation, University of
Portsmouth, Portsmouth~PO1~2EG, UK}

\author{Takashi Hiramatsu}
\email{hiramatsu.AT.resceu.s.u-tokyo.ac.jp} \affiliation{Research
Center for the Early Universe (RESCEU), School of Science,
University of Tokyo, 7-3-1 Hongo, Bunkyo, Tokyo 113-0033, Japan}

\author{Kazuya Koyama}
\email{kazuya.koyama.AT.port.ac.uk}%
\affiliation{Institute of Cosmology \& Gravitation, University of
Portsmouth, Portsmouth~PO1~2EG, UK}

\author{Sanjeev S.~Seahra}
\email{sanjeev.seahra.AT.port.ac.uk} %
\affiliation{Institute of Cosmology \& Gravitation, University of
Portsmouth, Portsmouth~PO1~2EG, UK}

\date{May 11, 2007}

\begin{abstract}

We study the behaviour of scalar perturbations in the
radiation-dominated era of Randall-Sundrum braneworld cosmology by
numerically solving the coupled bulk and brane master wave
equations. We find that density perturbations with wavelengths less
than a critical value (set by the bulk curvature length) are
amplified during horizon re-entry. This means that the radiation era
matter power spectrum will be at least an order of magnitude larger
than the predictions of general relativity (GR) on small scales.
Conversely, we explicitly confirm from simulations that the spectrum
is identical to GR on large scales.
%The comoving value of the critical scale is less than 10 AU for
% reasonable parameters,
Although this magnification is not relevant for the cosmic microwave
background or measurements of large scale structure, it
will have some bearing on the formation of primordial black holes
in Randall-Sundrum models.

\end{abstract}

\maketitle

\section{Introduction}\label{sec:introduction}

Our view of cosmology has been revolutionized by the notion that the
universe may be a lower-dimensional object embedded in some
higher-dimensional space.  A particularly simple realization of this
idea is furnished by the Randall-Sundrum (RS) braneworld model
\cite{Randall:1999vf}, which postulates that our observable universe
is a thin 4-dimensional hypersurface residing in 5-dimensional
anti-de Sitter (AdS) space (see
Refs.~\cite{Langlois:2002bb,Maartens:2003tw,Brax:2004xh} for
reviews).  Ordinary matter degrees of freedom are assumed to be
confined to the brane, while gravitational degrees of freedom are
allowed to propagate in the full 5-dimensional bulk. The warping of
AdS space allows us to recover ordinary general relativity (GR) at
distances greater than the curvature radius of the bulk $\ell$.
Current laboratory tests of Newton's law constrain $\ell$ to be less
than around $0.1\,\text{mm}$ \cite{Kapner:2006si}.

The cosmological implications of the Randall-Sundrum scenario have
been extensively studied.  It is well known that the Friedmann
equation governing the expansion of the brane universe differs from
general relativity by a correction of order $\rho/\sigma$, where
$\rho$ is the density of brane matter and $\sigma \gtrsim
(\text{TeV})^4$ is the brane tension.  The magnitude of this
correction defines the ``high-energy'' regime of braneworld
cosmology as the era when $\rho \gtrsim \sigma$ or equivalently
$H\ell \gtrsim 1$, where $H$ is the Hubble parameter.  At high
energies, the RS Friedmann equation implies $H \propto \rho$, which
results in dynamics significantly different from the standard $H^2
\propto \rho$ expansion law.

The major outstanding issue in RS cosmology is the behaviour of
cosmological perturbations \cite{Mukohyama:2000ui,Kodama:2000fa,
vandeBruck:2000ju, Koyama:2000cc, Langlois:2000ph}. The equations of
motion governing fluctuations of the model are found to differ from
GR in two principal ways at early times: First, they acquire
$\mathcal{O}(\rho/\sigma)$ high-energy corrections similar to those
found in the Friedmann equation.  By themselves, such corrections
are not difficult to deal with: they just modify the second-order
ordinary differential equations (ODEs) governing perturbations in
GR.  But the second type of modification is more problematic:
Perturbations on the brane are also coupled to fluctuations of the
5-dimensional bulk geometry, which are collectively known as the
``Kaluza Klein'' (KK) degrees of freedom of the model.  The KK modes
are governed by master partial differential equations (PDEs) defined
throughout the AdS bulk \cite{Mukohyama:2000ui,Kodama:2000fa}.  In
the original work of Randall \& Sundrum \cite{Randall:1999vf}, the
brane had a simple Minkowski geometry and the KK mode master wave
equations were solvable via separation of variables.  But the motion
of the brane in the cosmological case breaks the time-translation
symmetry of the bulk, which makes a simple separable solution
unattainable in most cases. The exception is the case of a brane
undergoing de Sitter inflation, where many analytic and
semi-analytic results are now available
\cite{Koyama:2004ap,Koyama:2005ek,Hiramatsu:2006cv,deRham:2007db,Koyama:2007as}.

Many authors have considered various schemes to solve the
perturbation problem without dealing with the KK master PDEs
directly. The most straightforward approach is to simply set the
KK degrees of freedom to zero while retaining
$\mathcal{O}(\rho/\sigma)$ corrections, which we refer to as the
``4-dimensional effective theory''. There are several
approximation methods that attempt to move beyond this simplest
effective theory in the case of tensor and scalar perturbations
\cite{Gordon:2000pt,Langlois:2000iu,Brax:2001qd,Bridgman:2001mc,
Koyama:2001ct,Leong:2001qm,%
Leong:2002hs,Easther:2003re,Battye:2003ks,Kobayashi:2004wy,Battye:2004qw,%
Koyama:2004cf}.  These generally consider the behaviour of
fluctuations during certain limiting regimes (i.e, high-energy or
low-energy), but the only known way of tackling the problem on all
scales simultaneously is by direct numerical solution of the
equations of motion.

The problem of tensor perturbations is somewhat simpler than the
scalar case.  The reason is that in linear theory, tensor modes
are pure gravitational degrees of freedom that do not couple to
matter fluctuations (provided that the matter anisotopic stress is
neglected). In the braneworld context, this means that there are
no brane-confined tensor degrees of freedom. The perturbation
problem formally reduces to solving a wave equation in the bulk
with boundary conditions imposed on a moving brane. Numeric
solutions
have been obtained by several groups %
\cite{Hiramatsu:2003iz,Hiramatsu:2004aa,Hiramatsu:2006bd,Kobayashi:2005dd,Kobayashi:2005jx,%
Kobayashi:2006pe,Ichiki:2003hf,Seahra:2006tm}. Interestingly, the
effective theory predicts a blue spectrum (i.e.~excess power) in
the stochastic gravitational wave background at frequencies
corresponding to modes that enter the Hubble horizon in the
high-energy regime. However, once KK effects are incorporated via
numeric simulations, one finds that the true spectrum is flat and
virtually identical to GR.  That is, the magnitudes of the
$\mathcal{O}(\rho/\sigma)$ and KK effects are nearly the same, but
they act in opposite sense and end up canceling each other out.
Physically, the reason that the KK modes cause a suppression of
the amplitude at high frequencies is that they efficiently radiate
gravitational wave energy from the brane into the bulk.

The purpose of the paper is to numerically solve for the behaviour
of scalar perturbations in the radiation-dominated regime of
braneworld cosmology. (Numeric analysis of the scalar perturbations
during inflation has already been done in
Refs.~\cite{Hiramatsu:2006cv,Koyama:2007as}.)  Unlike the tensor
case, there are several scalar degrees of freedom residing on the
brane, such as the density contrast $\delta\rho/\rho$.  The problem
reduces to the solution of a bulk wave equation coupled to a master
boundary field satisfying its own second order ODE on the moving
brane. Recently, two of the numerical codes developed to deal with
tensor perturbations \cite{Hiramatsu:2003iz,Seahra:2006tm} have been
generalized to handle boundary degrees of freedom
\cite{Hiramatsu:2006cv,Cardoso:2006nh}.  We use both codes in this
paper, which gives us the ability to confirm the consistency of our
numeric results via two independent algorithms.  We are ultimately
interested in finding the matter transfer function in the radiation
era, and also determining the relative influence of KK and
high-energy effects on the density perturbations. Heuristically, we
may expect the KK modes to amplify high-energy/small-scale density
perturbations, which is the opposite effect from the tensor case.
The reason is that we know that the gravitational force of
attraction in the RS model is stronger than in GR on scales less
than $\ell$ \cite{Randall:1999vf,Garriga:1999yh}.   This implies
that modes with a physical wavelength smaller than $\ell$ during
horizon crossing will be amplified due to the KK enhancement of the
gravitational force. However, this physical reasoning needs to be
tested with numeric simulations.

The layout of the paper is as follows:  The background geometry of
the RS cosmology we consider is described in \S\ref{sec:background}.
The formulae governing the gauge invariant scalar perturbations are
given in \S\ref{sec:scalar perturbations}.  Also in that section, we
derive new analytic approximations for various perturbation
variables at very high energies.  In \S\ref{sec:numeric}, we present
the numeric algorithms that we use to solve the coupled system of
bulk and brane wave equations.  Our results for the matter transfer
function are also contained in that section.  Finally, we summarize
and discuss the implications of our work in \S\ref{sec:discussion}

\section{Background geometry and brane dynamics}\label{sec:background}

As discussed in the Introduction, the Randall-Sundrum model we will
be considering consists of a thin brane, which is realized as
singular 4-surface, residing in 5-dimensional anti-de Sitter space.
In Poincar\'e coordinates, the bulk metric is given by the line
element
\begin{equation}\label{eq:5D metric 1}
    ds_5^2 = \frac{\ell^2}{z^2} (-d\tau^2 + \delta_{ij} dx^i dx^j +dz^2).
\end{equation}
Here, $\ell$ is the curvature scale of the bulk.  The boundary
of this spacetime given by a brane is described parametrically by
\begin{equation}
    \tau = \tau_\b(\eta), \quad z = z_\b(\eta).
\end{equation}
The parameter $\eta$ is defined to be the conformal time along the
brane, which leads to the following induced line element
\begin{equation}
    d\eta^2 = d\tau_\b^2 - dz_\b^2, \quad ds_\b^2 = a^2(-d\eta^2 +
    d\mathbf{x}^2) = -dt^2 + a^2 d\mathbf{x}^2.
\end{equation}
Here, we have identified the scale factor as $a(\eta) =
\ell/z_\b(\eta)$ and $t$ as the cosmic time. Now, we define $u^a =
dx^a_\b/dt$ to be the 5-velocity field of comoving observers on the
brane, while $n^a$ is the brane normal. These vector fields can then
be used to construct directional derivatives parallel and orthogonal
to the brane:
\begin{subequations}
\begin{align}
    \di_u & = u^a \di_a = \frac{1}{a} \left(\sqrt{1+H^2\ell^2}
    \frac{\di}{\di \tau} -
    H\ell \frac{\di}{\di z} \right)= \frac{\di}{\di t} = \frac{1}{a}
    \frac{\di}{\di\eta}, \\ \di_n & = n^a
    \di_a = \frac{1}{a} \left( -H\ell \frac{\di}{\di \tau} +
    \sqrt{1+H^2\ell^2} \frac{\di}{\di z} \right),
\end{align}
\end{subequations}
respectively.  Here, $H$ is the usual Hubble parameter
\begin{equation}
    H = \left( \frac{\di_u r}{r} \right)_\b = \frac{1}{a} \frac{da}{dt},
    \quad \frac{r}{\ell} = \frac{\ell}{z},
\end{equation}
where we have introduced the alternative bulk coordinate $r$.  We
assume that the matter on the brane has a fluid-like stress-energy
tensor with density $\rho$ and pressure $p$:
\begin{equation}
    T_{ab} = (\rho+p) u_a u_b + p h_{ab}, \quad g_{ab} = h_{ab} + n_a
    n_b.
\end{equation}
The junction conditions are given in terms of the extrinsic
curvature $K_{ab}$ of the brane,
\begin{equation}\label{eq:extrinsic}
    K_{ab} = h_a{}^c \nabla_c n_b, \qquad 2K_{ab} = -\kappa_5^2
    [T_{ab} - \tfrac{1}{3} h_{ab} (T-\sigma) ],
\end{equation}
where $\kappa_5^2 = 1/M_5^3$ is the gravity-matter coupling in 5
dimensions and $\sigma$ is the brane tension.  This yields
\begin{equation}
    \left( \frac{\di_n r}{r} \right)_\b =
    -\frac{\kappa_5^2(\rho+\sigma)}{6} =
    -\frac{\sqrt{1+H^2\ell^2}}{\ell}.
\end{equation}
We assume the standard Randall-Sundrum fine-tuning
\begin{equation}
    \kappa_5^2 = \kappa_4^2 \ell = 8\pi G \ell = \frac{6}{\sigma\ell},
\end{equation}
to obtain the Friedmann equation
\begin{equation}\label{eq:Friedmann}
    H^2 = \frac{1}{\ell^2} \frac{\rho}{\sigma} \left(
    2 + \frac{\rho}{\sigma} \right) = \frac{8\pi G}{3} \rho \left(
    1 + \frac{\rho}{2\sigma} \right).
\end{equation}
Note the high-energy $\mathcal{O}(\rho/\sigma)$ correction to the
expansion rate.  Finally, note that we have the usual conservation
of stress-energy on the brane; i.e.,
\begin{equation}\label{eq:conservation}
    h^{ab} \nabla_a T_{bc} = 0, \qquad \frac{d\rho}{dt} = -3(1+w)\rho H, \qquad w =
    \frac{p}{\rho}.
\end{equation}

\section{Scalar perturbations}\label{sec:scalar perturbations}

\subsection{Gauge invariant bulk perturbations}\label{sec:bulk
perturbations}

We now turn our attention to the perturbation of the cosmology
introduced in \S\ref{sec:background}.  It has been shown in
Refs.~\cite{Mukohyama:2000ui,Kodama:2000fa} that scalar-type
perturbations of the bulk geometry (\ref{eq:5D metric 1}) are
governed by a single gauge invariant master variable $\Omega$.  We
summarize those results in this subsection.

First, we introduce a harmonic basis with mode functions\footnote{To
describe the perturbations in the RS model we mostly follow the
notation and conventions of \citet{Kodama:2000fa}, which are based
on the 4-dimensional formalism of \citet{Kodama:1985bj}.}
\begin{equation}
    Y = e^{i\mathbf{k} \cdot \mathbf{x}}, \qquad Y_i = -\frac{1}{k}
    \di_i Y, \qquad Y_{ij} = \frac{1}{k^2} \di_i \di_j Y +
    \frac{1}{3} \delta_{ij} Y.
\end{equation}
Using these mode functions, we write the perturbed bulk metric as
\begin{equation}\label{eq:pertubed bulk metric}
    ds_5^2 = (q_{\alpha\beta} + f_{\alpha\beta})dx^\alpha dx^\beta +
    \frac{2r}{\ell} f_\alpha Y_i
    dx^\alpha dx^i + \frac{r^2}{\ell^2} \left[ (1+2H_L)\delta_{ij} +
    2H_T Y_{ij} \right] dx^i dx^j,
\end{equation}
where
\begin{equation}
    q_{\alpha\beta} dx^\alpha dx^\beta = \frac{\ell^2}{z^2} (-d\tau^2 + dz^2).
\end{equation}
Implicit integration over the wavevector $\mathbf{k}$ is understood
in equations like (\ref{eq:pertubed bulk metric}). A priori, the
particular form of the perturbation variables $f_{\alpha\beta}$,
$f_\alpha$, $H_L$ and $H_T$ depend on the choice of gauge. However,
one can construct gauge invariant combinations as follows
\begin{equation}
    F = H_L + \frac{1}{3} H_T + \frac{X^\alpha D_\alpha r}{r}, \qquad
    F_{\alpha\beta} = f_{\alpha\beta} + D_\alpha X_\beta + D_\beta X_\alpha,
\end{equation}
where
\begin{equation}
    X_\alpha = \frac{r}{k\ell} \left( f_\alpha + \frac{r}{k\ell} D_\alpha H_T
    \right),
\end{equation}
and $D_\alpha$ is the covariant derivative associated with the
2-metric $q_{\alpha\beta}$. The gauge invariant quantities are then
given in terms of the master variable $\Omega = \Omega(t,z)$
\begin{equation}
    F = \frac{1}{6r} \left( D^\mu D_\mu - \frac{2}{\ell^2} \right)
    \Omega, \quad F_{\alpha\beta} = \frac{1}{r} \left[ D_\alpha D_\beta -
     \frac{1}{3} q_{\alpha\beta} \left(
    2 D^\mu D_\mu - \frac{1}{\ell^2} \right) \right]
    \Omega.
\end{equation}
The bulk master variable satisfies the following wave equation
\begin{equation}\label{eq:bulk wave equation}
    0 = -\frac{\di^2\Omega}{\di \tau^2} + \frac{\di^2\Omega}{\di z^2}
    + \frac{3}{z} \frac{\di\Omega}{\di z} + \left( \frac{1}{z^2} -
    k^2 \right) \Omega.
\end{equation}

\subsection{Gauge invariant brane perturbations}\label{sec:brane
perturbations}

In addition to perturbations of the bulk metric, we must also
consider perturbations of the brane metric.  These are
parameterized as
\begin{equation}
    ds_\b^2 = (1-2\alpha Y)dt^2 - 2a\beta Y_i dt \, dx^i +
    a^2 \left[ (1+2h_L)\delta_{ij} +
    2h_T Y_{ij} \right] dx^i dx^j.
\end{equation}
As in 4 dimensions, the gauge-dependant variables
$(\alpha,\beta,h_L,h_T)$ can be used to construct gauge-invariant
quantities
\begin{equation}
    \Phi = h_L + \frac{1}{3} h_T - \frac{Ha}{k} \sigma_g, \qquad
    \Psi = \alpha - \frac{1}{k} \frac{d}{dt} (a\sigma_g),
\end{equation}
where
\begin{equation}
    \sigma_g = \frac{a}{k} \frac{dh_T}{dt} - \beta.
\end{equation}

As demonstrated in Ref.~\cite{Shiromizu:1999wj}, the effective
Einstein equation on the brane is in general given by
\begin{equation}\label{eq:effective Einstein equation}
    {}^{(4)}G_{ab} = \kappa_4^2 T_{ab} + \kappa_5^2 \Pi_{ab} -
    \Weyl_{ab}, \qquad \Pi_{ab} =  -\frac{1}{4} T_{ac}T^{c}{}_b +
    \frac{1}{12} T T_{ab} +
    \frac{1}{8} T_{cd} T^{cd} h_{ab} - \frac{1}{24} T^2
    h_{ab}.
\end{equation}
Here, $\Weyl_{ab}$ represents the projection of the electric part
of the bulk Weyl tensor onto the brane.  Notice that $\Weyl_{ab}$
satisfies
\begin{equation}\label{eq:Weyl conservation}
    h^{ab} \Weyl_{ab} = 0, \qquad \nabla_a (\kappa_5^2 \Pi^{ab} - \Weyl^{ab} ) =
    0,
\end{equation}
and vanishes in the background geometry discussed in
\S\ref{sec:background}. When perturbing (\ref{eq:effective
Einstein equation}), one can treat $\Weyl_{ab}$ as an additional
fluid source with a radiation-like equation of state.  Hence, we
parameterize the perturbations of $T_{ab}$ and $\Weyl_{ab}$ as
\begin{subequations}
\begin{align}
    \delta T_0{}^0 & = -\delta\rho \, Y, & \delta
    \Weyl_{0}{}^0 & = \kappa_4^2 \delta\rho_\Weyl Y, \\
    \delta T_i {}^0 & = -k Y_i \delta q , & \delta
    \Weyl_i{}^0 & = \kappa_4^2 k Y_i \delta q_\Weyl, \\
    \delta T_i{}^j & = \delta p  Y \delta_i{}^j +
    k^2 \delta\pi Y_i{}^j, & \delta \Weyl_i{}^j & =
    -\kappa_4^2 (\tfrac{1}{3} \delta\rho_\Weyl
    Y \delta_i{}^j + k^2 \delta\pi_\Weyl
    Y_i{}^j).
\end{align}
\end{subequations}
Henceforth, we will assume that the matter anisotropic stress
vanishes $\delta\pi = 0$.  The Weyl fluid perturbations
$(\delta\rho_\Weyl,\delta q_\Weyl,\delta \pi_\Weyl)$ are the
``Kaluza-Klein'' (KK) degrees of freedom of the model alluded
in the Introduction, since they represent the effects of bulk
geometry fluctuations on the brane.  The matter perturbation
variables above are not gauge invariant, but the following
quantities are:
\begin{subequations}
\begin{align}
    \rho \Delta & = \delta\rho - 3H\delta q, & a(\rho+p)V & =
     -k \delta q - (\rho + p) a \sigma_g, & \Gamma & = \delta
    p - c_s^2 \delta\rho, \\ \rho \Delta_\Weyl & = \delta\rho_\Weyl
    - 3H\delta q_\Weyl, &
    a(\rho+p)V_\Weyl & = -k \delta q_\Weyl - (\rho + p) a \sigma_g.
\end{align}
\end{subequations}
Here, we have defined the sound speed as $c_s^2 = \delta p /\delta
\rho$.  Since the Weyl fluid has the equation of state $\delta
p_\Weyl = \tfrac{1}{3}\delta\rho_\Weyl$, there is no KK entropy
perturbation $\Gamma_\Weyl = 0$. Furthermore, the KK anisotropic
stress is automatically gauge invariant.  Armed with these
definitions, it is possible to derive the following
gauge-invariant equations from the perturbation of
Eq.~(\ref{eq:effective Einstein equation}):
\begin{subequations}
\begin{align}
    \Phi & = \frac{4 \pi G \rho a^2}{k^2} \left[ \left(1 + \frac{\rho}
    {\sigma} \right) \Delta
    + \Delta_\Weyl \right], \\
    H\Psi - \frac{d\Phi}{dt} & = \frac{4\pi G (\rho+p) a}{k}
    \left[ \left(1 + \frac{\rho}{\sigma} \right) V + V_\Weyl \right],
    \\ \Phi + \Psi & = -8\pi G a^2 \delta\pi_\Weyl. \label{eq:KK anisotropic stress}
\end{align}
\end{subequations}
The first of these gives the high-energy $\mathcal{O}(\rho/\sigma)$
and KK corrections to the Poisson equation on the brane.  The second
equation gives the modifications of the standard equation governing
the evolution of the peculiar velocity field.  The third equation
demonstrates how the KK-modes can give rise to an anisotropic stress
on the brane.

Additional equations can be obtained from perturbing the equation
representing the conservation of matter stress energy on the
brane: $\delta(h^{ab}\nabla_a T_{bc})=0$.  These are
\begin{subequations}\label{eq:stress energy conservation
perturbed}
\begin{align}
    \frac{1}{a} \frac{d}{dt} (aV) & = \frac{k}{a}\Psi + \frac{k}{a}
    \frac{\Gamma+c_s^2\rho\Delta}{\rho(1+w)}, \\ \frac{1}{a^3}
    \frac{d}{dt} (a^3\rho\Delta) & = -\frac{k}{a}
    \rho(1+w)\left(1-\frac{3a^2}{k^2} \frac{dH}{dt} \right) V -
    3\rho(1+w)\left(\frac{d\Phi}{dt} - H\Psi\right).
\end{align}
\end{subequations}
Finally, we can also perturb the righthand expression in
Eq.~(\ref{eq:Weyl conservation}) to get two more equations, but
these are not sufficient to close the effective Einstein equations
(\ref{eq:effective Einstein equation}) on the brane.  Hence we can
only go so far with this effective Weyl fluid description; a
complete treatment must describe how the brane degrees of freedom
presented here are coupled to the bulk degree of freedom $\Omega$.

\subsection{Perturbation of the junction conditions}

The means to connect the bulk perturbations of \S\ref{sec:bulk
perturbations} and the brane perturbations of \S\ref{sec:brane
perturbations} is the perturbation of the junction conditions
(\ref{eq:extrinsic}).  This yields several results, the most
important of which is that $\Omega$ satisfies a boundary condition
on the brane
\begin{equation}\label{eq:boundary condition}
    \left[ \di_n \Omega + \frac{1}{\ell} \left(1 +
    \frac{\rho}{\sigma} \right) \Omega + \frac{6\rho a^3}{\sigma
    k^2} \Delta \right]_\b = 0.
\end{equation}
The perturbation of (\ref{eq:extrinsic}) also gives the KK gauge
invariants in terms of $\Omega$:
\begin{subequations}
\begin{align}
    \kappa_4^2 \rho \Delta_\Weyl & = \frac{k^2}{\ell a^3} \left[ \frac{k^2 + 3H^2 a^2}{3 a^2} \Omega
    - H \di_u \Omega \right]_\b, \\
    \kappa_4^2 a(\rho + p) V_\Weyl & = \frac{k^3}{3\ell a^2} \left[ H\Omega - \di_u \Omega
    \right]_\b, \\ \kappa_4^2 \delta\pi_\Weyl & = \frac{1}{2\ell a^3}
    \left[  \frac{1}{3} k^2 \Omega -H\di_u\Omega - \frac{3(\rho+p)}{\ell\sigma}
    \di_n \Omega + \di_u^2 \Omega \right]_\b.
\end{align}
\end{subequations}
These reproduce the results first derived in
Ref.~\cite{Deffayet:2002fn}.  Using these equations, we can find
explicit formulae for the gauge invariants defined above in terms of
$\Delta$ and $\Omega_\b(\eta) = \Omega(\tau_\b(\eta),z_\b(\eta))$:
\begin{subequations}\label{eq:explicit gauge invariants}
\begin{align}
    \label{eq:explicit Phi} \Phi & = \frac{3a^2\rho(\rho+\sigma)}{k^2 \ell^2 \sigma^2}
    \Delta + \left( \frac{3H^2 a^2+k^2}{6\ell a^3} \right) \Omega_\b
    - \frac{H}{2\ell a^2} \frac{d\Omega_\b}{d\eta}, \\ \label{eq:explicit Psi}  \Psi & =
    -\frac{3\rho a^2 (3w\rho+4\rho+\sigma)}{k^2\ell^2\sigma^2} \Delta -
    \left[ \frac{(3w+4)\rho^2}{2\ell^3 a\sigma^2} + \frac{(5+3w)\rho}{2\ell^3
    a\sigma} +
    \frac{k^2}{3\ell a^3} \right] \Omega_\b + \frac{3H}{2\ell a^2} \frac{d\Omega_\b}{d\eta}
    - \frac{1}{2\ell a^3} \frac{d^2\Omega_\b}{d\eta^2}, \\ \label{eq:explicit V} V & =
    \frac{3wHa}{k(1+w)} \Delta - \frac{1}{k(1+w)}
    \frac{d\Delta}{d\eta} + \frac{kH}{2\ell a^2} \Omega_\b -
    \frac{k}{2\ell a^3}
    \frac{d\Omega_\b}{d\eta}.
\end{align}
\end{subequations}
Other quantities of interest are the curvature perturbation on
uniform density slices,
\begin{equation}\label{eq:explicit zeta}
    \zeta =  \Phi - \frac{Ha V}{k} + \frac{\Delta}{3(1+w)} = \left[ \frac{1}{3}
    - \frac{3\rho a^2(w\sigma - \sigma -\rho)}{k^2\ell^2\sigma^2} \right]
    \frac{\Delta}{1+w} + \frac{Ha}{k^2(1+w)} \frac{d\Delta}{d\eta}
    + \frac{k^2}{6\ell a^3} \Omega_\b,
\end{equation}
and the curvature perturbation on comoving slices,
\begin{equation}
    \mathcal{R}_c =  \Phi - \frac{Ha V}{k} =
    - \frac{3\rho a^2(w\sigma - \sigma -\rho)}{k^2\ell^2\sigma^2}
    \frac{\Delta}{1+w} + \frac{Ha}{k^2(1+w)} \frac{d\Delta}{d\eta}
    + \frac{k^2}{6\ell a^3} \Omega_\b,
\end{equation}
Hence, if we have knowledge of the density contrast $\Delta$ and the
value of $\Omega$ on the brane, we can obtain the behaviour of all
the other gauge invariants via algebra and differentiation.

The above formulae can be easily manipulated to find a wave equation
for $\Delta$:
\begin{subequations}\label{eq:brane wave equation}
\begin{gather}
    \frac{d^2 \Delta}{d\eta^2} + (1+3c_s^2-6w) Ha \frac{d\Delta}{d\eta} +
    \left[c_s^2 k^2 + \frac{3\rho a^2}{\sigma\ell^2} A +
    \frac{3\rho^2 a^2}{\sigma^2\ell^2} B \right] \Delta = -\frac{k^2\Gamma}{\rho} + \frac{k^4(1+w)\Omega_\b}{3\ell a^3}, \\
    A = 6c_s^2 -1 -8w+3w^2, \qquad
    B = 3c_s^2-9w-4.
\end{gather}
\end{subequations}
The above ODE, the bulk wave equation (\ref{eq:bulk wave equation})
and the boundary condition (\ref{eq:boundary condition}) comprise a
closed set of equations for $\Delta$ and $\Omega$.

Before moving on, we have three remarks:  First, note that in the
low energy universe, we can neglect $\mathcal{O}(\rho^2/\sigma^2)$
terms. Then, the $\Delta$ wave equation reduces to
\begin{equation}
    \frac{d^2 \Delta}{d\eta^2} + (1+3c_s^2-6w) Ha \frac{d\Delta}{d\eta} +
    \left[c_s^2 k^2 + 4\pi G\rho a^2 (6c_s^2 - 1 - 8w +3w^2)
    \right] \Delta \approx -\frac{k^2\Gamma}{\rho} + \frac{k^4(1+w)\Omega_\b}{3\ell
    a^3}.
\end{equation}
Setting $\Omega_\b = 0$ yields the standard 4-dimensional
dynamical equation for $\Delta$; hence, we recover GR at low
energies. Second, we note that from the conservation of
stress-energy on the brane (\ref{eq:stress energy conservation
perturbed}), we must have
\begin{equation}\label{eq:zeta conservation}
    \frac{d\zeta}{d\eta} = -\frac{kV}{3} -
    \frac{Ha\Gamma}{\rho(1+w)}.
\end{equation}
We can easily verify that Eqns.~(\ref{eq:explicit V}),
(\ref{eq:explicit zeta}) and (\ref{eq:brane wave equation}) imply
that (\ref{eq:zeta conservation}) is satisfied identically, which
is a good consistency check of our formulae.  Finally, the entropy
perturbation can be simplified when the brane matter is either a
perfect fluid or a scalar field:
\begin{equation}
    \Gamma = (\Upsilon - c_s^2) \rho \Delta, \quad \Upsilon = \begin{cases}
    c_s^2, & \text{perfect fluid}, \\
    1, & \text{scalar field}. \end{cases}
\end{equation}
For the rest of this paper, we will be considering the perfect
fluid case.

\subsection{Asymptotic behaviour in the high-energy radiation-dominated
regime}\label{sec:asymptotic}

To model the perturbations in the very early universe, we assume
radiation domination and $\rho \gg \sigma$; or, equivalently, $H\ell
\gg 1$.  Under these circumstances, the following approximations
hold:
\begin{equation}\label{eq:high-energy approximations}
    w = 1/3, \quad a \approx a_0 (k\eta)^{1/3}, \quad \rho \approx \sigma H\ell,
    \quad \di_n \approx -\di_u = -\di_t.
\end{equation}
The last relationship can be used to re-cast the boundary condition
on the bulk master variable $\Omega$ into a 1st order ordinary
differential equation for $\Omega_\b$.  Hence, the equation of
motion for $\Delta$ and boundary condition become
\begin{equation}
    0 \approx \frac{d^2\Delta}{d\eta^2} + \left( \frac{k^2}{3} - \frac{4a_0 k^{1/3}}{3\ell \eta^{2/3}} -
    \frac{2}{\eta^2} \right) \Delta - \frac{4k^3}{9\ell a_0^3\eta}
    \Omega_\b, \quad \frac{d\Omega_\b}{d\eta} \approx \frac{1}{3\eta} \left( 1 + \frac{3a_0
    k^{1/3} \eta^{4/3}}{\ell} \right) \Omega_\b + \frac{2\ell a_0^3}{k} \Delta.
\end{equation}
By differentiating the first equation we can derive a decoupled
third order equation for $\Delta$, which can be solved via power
series methods.  The solutions for $\Delta$ and $\Omega$ are a
superposition of three linearly independent modes:
\begin{equation}
    \Delta = \sum_{i=1}^3 \mathcal{A}_i \Delta^{(i)}, \quad
    \Omega_\b = \sum_{i=1}^3 \mathcal{A}_i \Omega_\b^{(i)},
\end{equation}
where the $\mathcal{A}_i$ are constants.  At leading order in
$(k\eta)$, these mode functions are given by
\begin{subequations}\label{eq:high-energy solutions}
\begin{align}
    \text{dominant growing mode:} & &  \Delta^{(1)} & \approx \tfrac{4}{3} (k\eta)^2
    , &
    \Omega_\b^{(1)} & \approx a_0^3 k^{-2} \ell (k\eta)^3
    , \\
    \text{subdominant growing mode:} & &
    \Delta^{(2)} & \approx (k\eta)^{4/3},
    & \Omega_\b^{(2)} & \approx -\tfrac{7}{2} a_0^3 k^{-2} \ell (k\eta)^{1/3}
    ,\\
    \text{decaying mode:} & & \Delta^{(3)} & \approx \tfrac{10}{3}
    k\ell a_0^{-1} (k\eta)^{-1}, &
    \Omega_\b^{(3)} & \approx -20 a_0^2 k^{-1} \ell^2.
\end{align}
\end{subequations}
As the labels on the left suggest, the growth of density
perturbations for the dominant growing mode is faster than for the
subdominant growing mode on superhorizon scales.  The density
contrast of the third mode decays in the high-energy regime.

We can calculate the behaviour of $\zeta$ and $\Psi$ for the
dominant growing mode by substituting the full series solutions
for $\Delta^{(1)}$ and $\Omega^{(1)}$ into (\ref{eq:explicit Psi})
and (\ref{eq:explicit zeta}), which yields:
\begin{equation}\label{eq:zeta Psi approx}
    \zeta^{(1)} \approx 1, \qquad \Psi^{(1)} \approx -4 \zeta^{(1)}.
\end{equation}
We see that the dominant growing mode curvature perturbation and
Newtonian potential are conserved on superhorizon scales.  Also
note that (\ref{eq:zeta Psi approx}) implies
\begin{equation}\label{eq:Delta approx}
    \Delta^{(1)} = \frac{4}{27}\left( \frac{k}{Ha} \right)^2 \zeta^{(1)}.
\end{equation}
Eqs.~(\ref{eq:zeta Psi approx}) and (\ref{eq:Delta approx}) are
similar to the standard superhorizon growing mode results for a
radiation dominated universe in general relativity,
\begin{equation}\label{eq:GR relations}
    \Psi^{(1)}_\GR \approx -\Phi^{(1)}_\GR \approx -\frac{3(1+w)}
    {3w+5} \zeta^{(1)}_\GR =  -\frac{2}{3} \zeta^{(1)}_\GR, \quad
    \Delta_\GR^{(1)} \approx \frac{(3w+1)^2(1+w)}{2(3w+5)} \left(
    \frac{k}{Ha} \right)^2 \zeta^{(1)}_\GR =
    \frac{4}{9}\left( \frac{k}{Ha} \right)^2
    \zeta^{(1)}_\GR,
\end{equation}
but the numerical factors are different.\footnote{It is
interesting to note that it is impossible to find an effective
equation of state parameter $w_\eff$ to make the general GR
formulae given in (\ref{eq:GR relations}) compatible with
(\ref{eq:zeta Psi approx}) and (\ref{eq:Delta approx}). Some
authors \cite{Sendouda:2006nu} have previously tried to describe
the high-energy radiation epoch of RS cosmology with an effective
fluid with $w_\eff = 5/3$, but we see that this would predict
$\Delta^{(1)} \approx 24/5 (k/Ha)^2 \zeta^{(1)}$ and $\Psi^{(1)}
\approx -4 \zeta^{(1)}/5$, which is in clear conflict with the
correct results (\ref{eq:zeta Psi approx}) and (\ref{eq:Delta
approx}).
%KK
This is due to a large Weyl anisotropic stress that modifies the GR
relationship between $\Psi$ and $\Phi$. }

Finally, we note that the method described in the subsection fails
to yield an approximate solution for the other metric potential
$\Phi^{(1)}$.  The reason is that when the full series expansions
for $\Delta^{(1)}$ and $\Omega_\b^{(1)}$ are substituted into
(\ref{eq:explicit Phi}), the leading order contribution vanishes,
leaving a result that is the same order as the error in the original
approximations (\ref{eq:high-energy approximations}).  We must
therefore rely on numeric simulations to determine the asymptotic
behaviour of $\Phi$.

\section{Numeric analysis}\label{sec:numeric}

\subsection{Dimensionless parameters and integration algorithms}

Our goal in the section is the solution of the system comprised of
the bulk wave equation (\ref{eq:bulk wave equation}) and the brane
wave equation (\ref{eq:brane wave equation}) subject to the boundary
condition (\ref{eq:boundary condition}).  For the rest of the paper,
we will restrict ourselves to the case of a radiation-dominated
brane with $w = 1/3$.  To perform the analysis, it is necessary to
make the various quantities in the equations dimensionless.  To do
so, we define the ``*'' epoch as the moment in time when a mode with
wavenumber $k$ enters the Hubble horizon
\begin{equation}
    k = H_* a_*, \qquad H_* = H(\eta_*), \qquad a_* = a(\eta_*),
    \qquad z_* = z_\b(\eta_*).
\end{equation}
Then, we introduce dimensionless/normalized quantities as
\begin{equation}
    \hat\Omega = \frac{\Omega}{a_*\ell^3}, \qquad \rhohat =
    \frac{\rho}{\sigma}, \qquad \H = H\ell, \qquad \k = \H_* = \frac{k\ell}{a_*}
    = kz_* =
    \sqrt{\rhohat_*(\rhohat_* + 2)}, \qquad \a = \frac{a}{a_*}.
\end{equation}
Another important era is the critical epoch when $\H_c = H_c\ell =
1$ and the radiation density has its critical value $\rhohat_c =
\sqrt{2}-1$.  We define the critical epoch as the transition between
high and low energy regimes. The ratio of the wavenumber of any
given mode to the wavenumber $k_\c = H_\c a_\c$ of the critical mode
that enters the horizon at the critical epoch is
\begin{equation}
    \frac{k}{k_\c} = \sqrt{\rhohat_*(\rhohat_*+2)} \left(
    \frac{\sqrt{2}-1}{\rhohat_*} \right)^{1/4},
\end{equation}
where $a_\c$ is the value of the scale factor at the critical
epoch. Generally speaking, we call modes with $k > k_\c$
``supercritical'' and modes with $k<k_\c$ ``subcritical''.  The
scale defined by the critical mode in today's universe (with scale
factor $a_0$) is given by
\begin{equation}\label{eq:scale today}
    \frac{a_0}{k_\c} = 1.4 \times
    10^{12} \left( \frac{\ell}{0.1\,\text{mm}} \right)^{1/2} \left( \frac{g_\c}{100}
    \right)^{1/12} \text{m},
\end{equation}
where $g_\c$ is the number of relativistic degrees of freedom in the
matter sector at the critical epoch.  For $\ell = 0.1\,\text{mm}$,
this corresponds to a scale of $\sim 10$ astronomical units (AU),
which is incredibly tiny by cosmological standards.  Finally, if we
normalize all coordinates by $z_*$,
\begin{equation}
    \hat\tau = \tau/z_*, \quad \hat z = z/z_*, \quad \hat\eta =
    \eta/z_*,
\end{equation}
we find that the entire system of master equations is characterized
by the dimensionless matter density at horizon crossing $\rhohat_*$.

\begin{figure}
\begin{center}
    \includegraphics{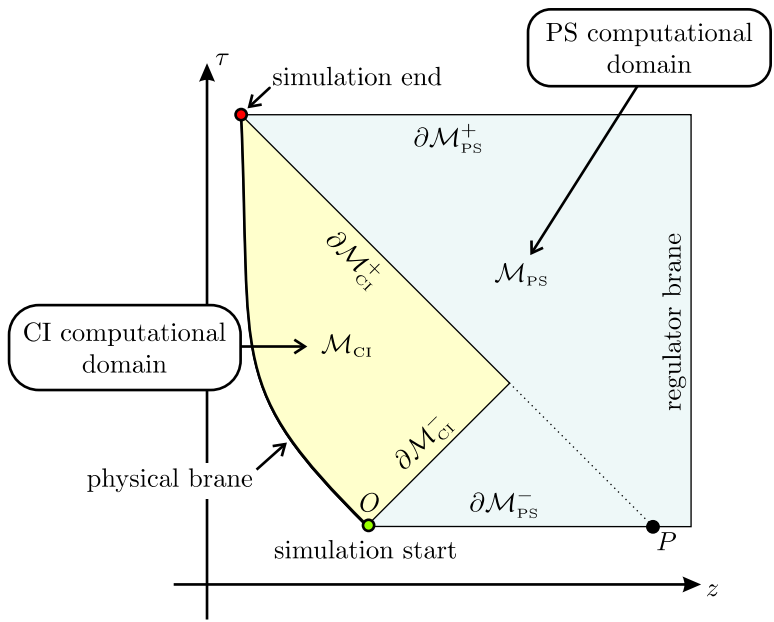} \caption{The
    computational domains employed by the pseudo-spectral (PS) and
    characteristic integration (CI) methods.  The future and past
    boundaries $\di\mathcal{M}_\PS^\pm$ of the PS domain $\mathcal{M}_\PS$
    are lines of constant $\tau$, while for the CI method they are null
    rays.  In the PS method, one needs a constant $z$ regulator brane to
    render the computational domain finite.  In principle, the position of
    the regulator is free, but it should be placed to the right of the
    point $P$ in order to be outside the causal past of the physical brane.
    Initial conditions for the two methods are placed on
    $\di\mathcal{M}_\PS^-$ and $\di\mathcal{M}_\CI^-$,
    respectively.\label{fig:compuational domains}}
\end{center}
\end{figure}

Once the system has been reduced into the dimensionless form, we
have two independent numerical codes that can be used to solve for
$\Delta$ and $\hat\Omega_\b$.  The first is the pseudo-spectral (PS)
method used in Ref.~\cite{Hiramatsu:2006cv, Hiramatsu:2006bd}
and the second is the characteristic integration (CI) algorithm developed in
Ref.~\cite{Cardoso:2006nh}\footnote{To apply the CI method as
described in \cite{Cardoso:2006nh}, the bulk wave equation needs to
be mapped into a canonical form via the change of variable
$\hat\Omega = (z_*/z)^{3/2}\phi$.}; detailed descriptions of each
method can be found in the respective papers. There is one technical
difference between the two algorithms that is worth highlighting
here:  Namely, each code solves for $\hat\Omega$ in different
regions of the 5-dimensional spacetime, as shown in
Fig.~\ref{fig:compuational domains}.  One implication of this is
that the initial data for each code needs to be specified at
different places.  For the PS code, one needs to choose $\hat\Omega$
and $\di\hat\Omega/\di\hat\tau$ on an initial spacelike hypersurface
$\di\mathcal{M}^-_\PS$ of constant $\hat\tau$, while for the CI
algorithm one needs $\hat\Omega$ only on an initial null
hypersurface $\di\mathcal{M}^-_\CI$ intersecting the brane.

What initial conditions should we actually use?  The calculations of
\S\ref{sec:asymptotic} suggest that at sufficiently early times, the
dynamics of the system are well described by three distinct modes
(\ref{eq:high-energy solutions}).  If our initial data surface is
positioned within this early era, it follows that generic choices of
initial conditions will excite some superposition of these three
modes. However, after a short period of time the dominant growing
mode will overtake the contributions from the subdominant growing
and decaying modes.  Hence, we expect that the late time dynamics of
our model will be insensitive to the choice of initial conditions,
provided that the initial data hypersurface is oriented at an early
enough time.  Since we do not expect the initial conditions to
matter very much, we make the simple choices
\begin{subequations}\label{eq:initial conditions}
\begin{align}
    \text{PS initial conditions: } & & \Delta(O) & = \mathcal{N} \a_i^6,
    & \hat\di_\eta \Delta(O) & = 6\mathcal{N}\H_i\a_i^7, & \hat\Omega(\di\mathcal{M}^-_\PS) & =
    0, & \hat\di_\tau \hat\Omega (\di\mathcal{M}^-_\PS) & = 6 \mathcal{N}\rhohat_* \a_i^6 /
    \k^2 \H_i,
    \\ \text{CI initial conditions: } & & \Delta(O) & = \mathcal{N}\a_i^6,
    & \hat\di_\eta \Delta(O) & = 6\mathcal{N}\H_i\a_i^7, & \hat\Omega(\di\mathcal{M}^-_\CI) & =
    0, & &
\end{align}
\end{subequations}
where $\hat\di_\eta = \di/\di\etahat$, $\hat\di_\tau =
\di/\di\hat\tau$, and an ``$i$'' subscript denotes the initial
value of the scale factor and Hubble parameter.  Here,
$\mathcal{N}$ is a normalization constant that we will often
select to make $\zeta = 1$ at the initial time.  These initial
conditions are motivated by the fact that we expect $\Delta
\propto \a^6 \gg \hat\Omega_\b$ for the dominant growing mode at
very early times, which can certainly be satisfied by setting
$\hat\Omega = 0$ on the initial data surface. For the PS method,
the initial time derivative of $\hat\Omega$ is selected to satisfy
the boundary condition (\ref{eq:boundary condition}) at the
initial time.  Note that the initial conditions for the two
methods are actually incompatible due to the different locations
of the initial surface. But as we have argued above and will see
below, this difference should make no difference to the final
results.  (We will test this assumption in \S\ref{sec:initial
data} below.)

\subsection{Typical waveforms}

\begin{figure}
\begin{center}
    \includegraphics{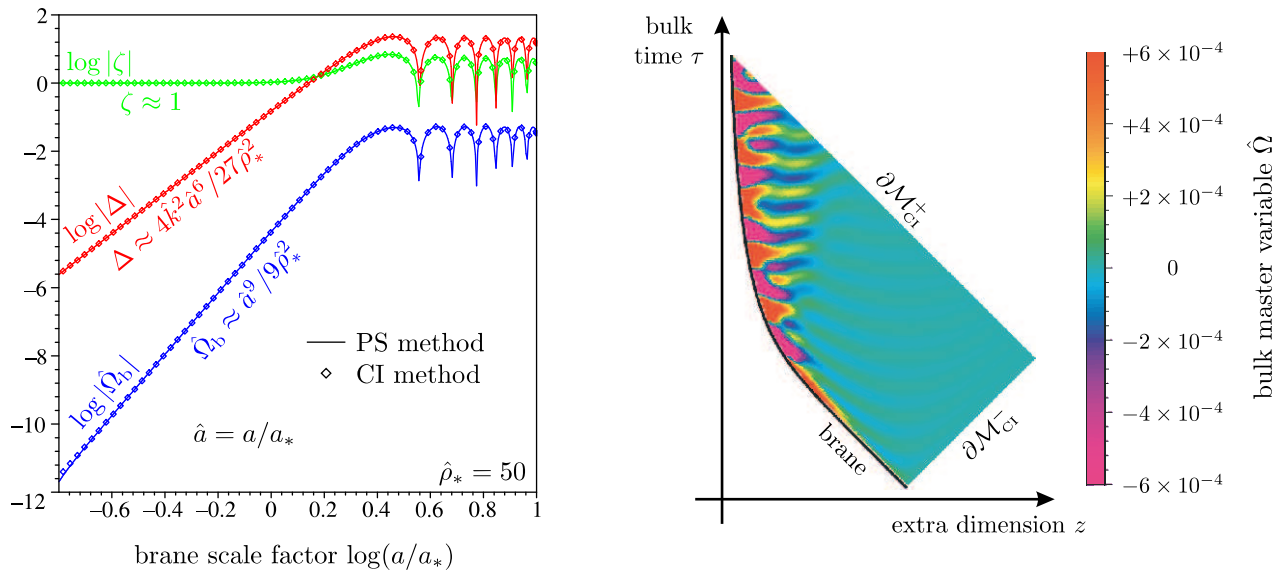}
    \caption{Comparison between typical results of the PS and CI
    codes for various brane quantities (\emph{left}); and the
    typical behaviour of the bulk master variable (\emph{right})
    as calculated by the CI method.  Very good agreement between
    the two different numerical schemes is seen in the left panel,
    despite the fact that they use different
    initial conditions.  We also see excellent consistency between the
    simulations and the approximations developed in
    \S\ref{sec:asymptotic} for the behaviour of the system on
    high-energy/superhorizon scales
    [c.f.~Eqs.~(\ref{eq:high-energy solutions})--(\ref{eq:Delta approx})].
    Also note that on subhorizon scales, $\Delta$ and $\zeta$
    undergo simple harmonic oscillations, which is consistent with the behaviour in
    GR (\ref{eq:GR Delta zeta}).  The bulk profile demonstrates
    our choice of initial conditions:  We see that the bulk master variable $\hat\Omega$
    is essentially zero during the early stages of the simulation,
    and only becomes ``large'' when the mode crosses the horizon.
    \label{fig:compare}}
\end{center}
\end{figure}

We now turn to the results of our simulations.  In
Fig.~\ref{fig:compare} we show the output of the PS and CI codes
for a typical simulation of a supercritical mode with $\rhohat_* =
50$. As expected we have excellent agreement between the two
codes, despite the fact that they use different initial conditions
(\ref{eq:initial conditions}). In fact, we have confirmed that the
codes give essentially identical results for a wide range of
parameters, which gives us confidence in our numerical methods.
It is also reassuring that the simulation results closely match
the analytic predictions of \S\ref{sec:asymptotic} on
high-energy/superhorizon scales.  Note that for all simulations,
we recover that $\Delta$ and $\zeta$ are phase-locked plane waves,
\begin{equation}
    \Delta(\eta) \propto \cos \frac{k\eta}{\sqrt{3}}, \quad \Delta(\eta)
    \approx 4 \zeta(\eta),
\end{equation}
at sufficiently late times $k\eta \gg 1$, which is actually the same
behaviour as seen in GR, where we have the following exact
solutions for the growing mode density contrast and curvature
perturbation during radiation domination:
\begin{subequations}\label{eq:GR Delta zeta}
\begin{align}
    \Delta^{(1)}_\GR & = \mathcal{A} \left( \frac{\sqrt{3}}{k\eta}\sin \frac{k\eta}{\sqrt{3}} -
    4 \cos \frac{k\eta}{\sqrt{3}} \right) \approx \mathcal{A} \begin{cases} \frac{4}{9}
    k^2\eta^2, & k\eta \ll 1, \\ -4 \cos \frac{k\eta}{\sqrt{3}}, &
    k\eta \gg 1, \end{cases} \\
    \zeta^{(1)}_\GR & = \mathcal{A} \left( \frac{2\sqrt{3}}{k\eta}\sin \frac{k\eta}{\sqrt{3}} -
    \cos \frac{k\eta}{\sqrt{3}} \right) \approx \mathcal{A}
    \begin{cases} 1, & k\eta \ll 1, \\ -\cos \frac{k\eta}{\sqrt{3}}, &
    k\eta \gg 1, \end{cases}
\end{align}
\end{subequations}
where $\mathcal{A}$ is a constant.
\begin{figure}
\begin{center}
    \includegraphics{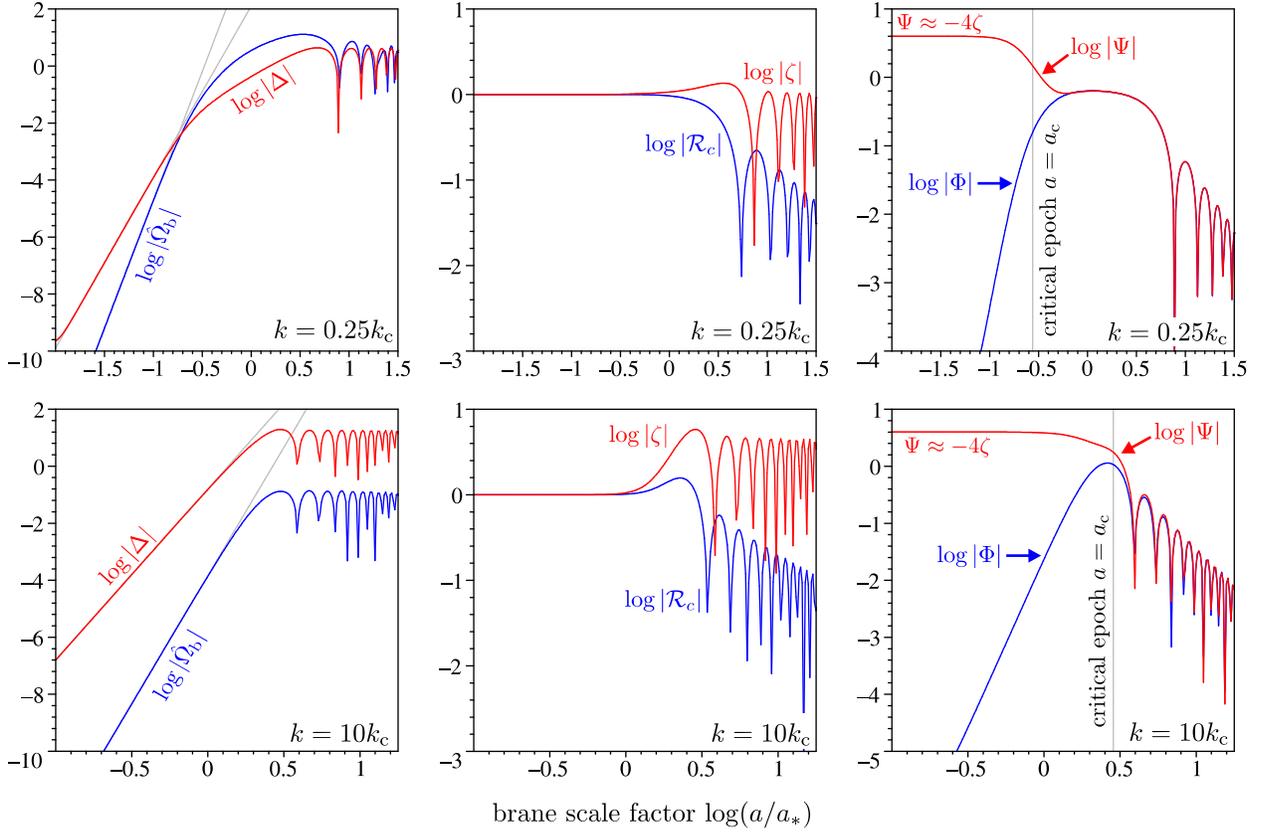} \caption{Typical behaviour of
    various brane gauge invariants for modes entering the horizon
    in subcritical (\emph{upper panels}) and supercritical
    (\emph{lower panels}) epochs.  For the plots on the left, the
    grey straight lines give the analytic expectations
    (\ref{eq:high-energy solutions}) for the behaviour of $\Delta$ and
    $\hat\Omega_\b$ on large scales and high energies.  The central plots show the
    comoving and uniform density curvature perturbations,
    $\mathcal{R}_c$ and $\zeta$ respectively.  As in conventional
    cosmology, these are conserved and equal on superhorizon
    scales.  Also note that the amplitude of $\zeta$ after horizon
    crossing is larger for the mode with larger $k$.
    The plots on the right show the two gauge invariant metric
    perturbations on the brane.  Before the critical epoch we have
    that $|\Psi| \gg |\Phi|$, while afterwards we recover the GR
    result $\Phi \approx -\Psi$. Finally, in the high-energy regime
    we see $\Psi \approx -4\zeta$, in line with the approximation
    (\ref{eq:zeta Psi approx}). \label{fig:typical}}
\end{center}
\end{figure}

In Fig.~\ref{fig:typical}, we show the simulated behaviour of
several different gauge invariants on the brane for given
subcritical ($k = 0.25 k_\c$) and supercritical ($k = 10 k_\c$)
modes. These plots illustrate several points that are generic to
all of our simulations:  The curvature perturbations
$\mathcal{R}_c$ and $\zeta$ are conserved on superhorizon scales
for all cases, just as in general relativity.  This useful fact
allows us to define the primordial amplitude of a given
perturbation as the value of $\zeta$ or $\mathcal{R}_c$ before
horizon crossing.  For the metric perturbations $\Phi$ and $\Psi$,
we find that for supercritical and superhorizon scales
\begin{equation}
    |\Psi| \gg |\Phi|, \quad \Psi \approx -4\zeta = \text{constant},
    \quad (\text{for $a \lesssim a_\c$ and $k \lesssim Ha$}).
\end{equation}
This implies that the Weyl anisotropic stress
$\kappa_4^2\delta\pi_\KK \approx -\Psi/a^2$ is relatively large in
the high-energy superhorizon regime.

\begin{figure}
\begin{center}
    \includegraphics{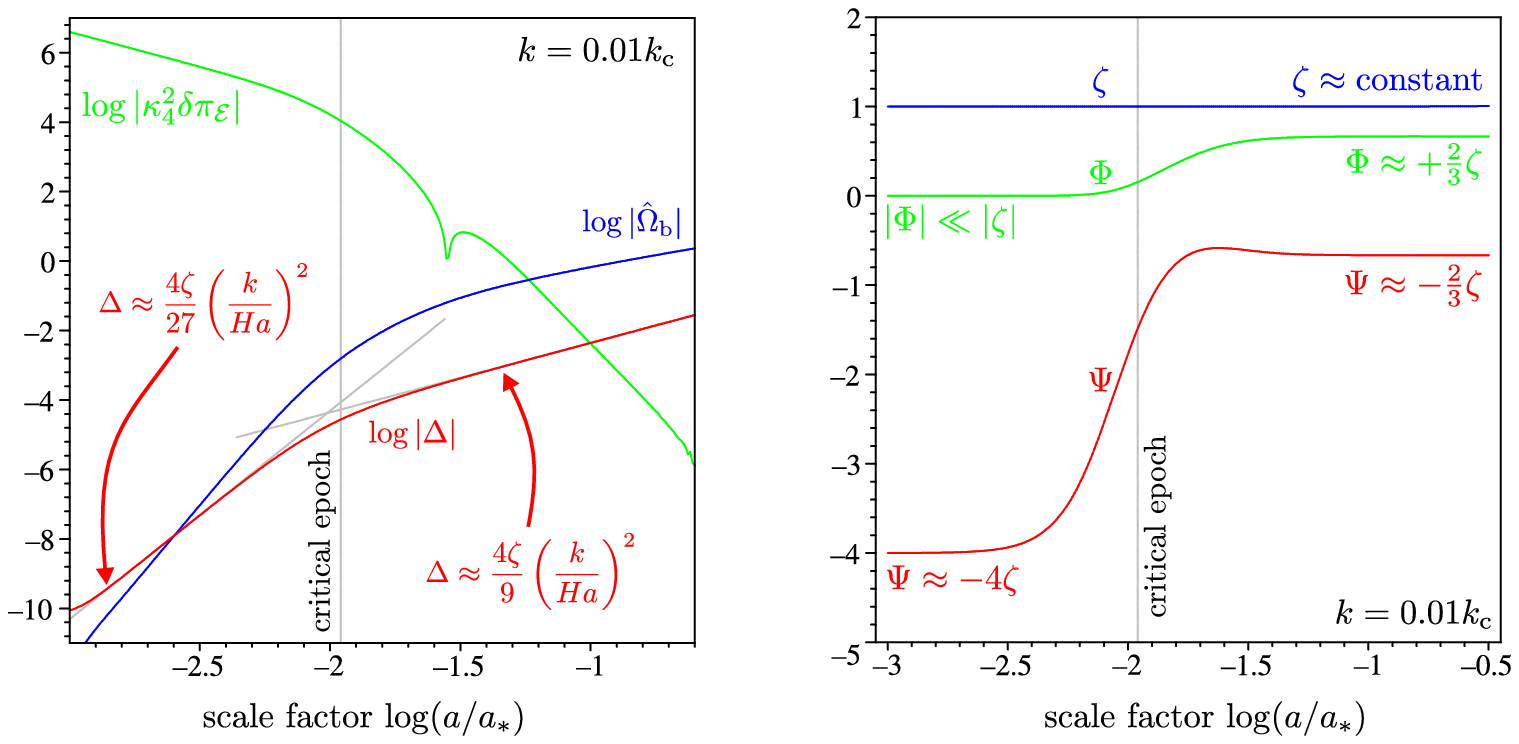} \caption{The simulated behaviour of
    an extremely subcritical $k \ll k_\c$ mode on superhorizon scales.  On the left we show
    how the $\Delta$ gauge invariant switches from the high-energy behaviour
    predicted in Eq.~(\ref{eq:Delta approx}) to the familiar GR result
    $(\ref{eq:GR relations})$ as the universe expands through the
    critical epoch.  We also show how the KK anisotropic stress $\kappa_4^2 \delta\pi_\KK$
    steadily decays throughout the simulation, which is typical of
    all the cases we have investigated.  On the right, we show the
    metric perturbations $\Phi$ and $\Psi$ as well as the curvature perturbation
    $\zeta$.  Again, note how the GR result $\Phi \approx -\Psi \approx -2\zeta/3$
    is recovered at low energy. \label{fig:subcritical}}
\end{center}
\end{figure}

Finally, Fig.~\ref{fig:subcritical} illustrates how the ordinary
superhorizon behaviour of perturbations in GR is recovered for
modes entering the Hubble horizon in the low energy era.  We see
how $\Delta$, $\Phi$ and $\Psi$ smoothly interpolate between the
non-standard high-energy behaviour to the usual expectations given
by (\ref{eq:GR relations}).  Also shown in this plot is the
behaviour of the KK anisotropic stress, which steadily decays
throughout the simulation.  This is characteristic of all the
cases we have studied.

\subsection{Enhancement factors and the transfer
function}\label{sec:transfer}

If we examine the curvature perturbations plots in
Fig.~\ref{fig:typical} in detail, we see that the amplitude of
$\zeta$ increases during horizon crossing.  Furthermore, the degree
of enhancement seems to increase with increasing $k$.  This is quite
different from the behaviour of the perturbations of a radiation
dominated universe in ordinary GR (\ref{eq:GR Delta
zeta}), where the asymptotic amplitude of $\zeta$ is the same before and
after horizon crossing.  Hence, in the braneworld case we see an
enhancement in the amplitude of perturbations that is not present in
conventional theory.

What is responsible for this enhancement?  As discussed in the
Introduction, there are actually two separate effects to consider:
First, there is the modification of the universe's expansion at high
energies and the $\mathcal{O}(\rho/\sigma)$ corrections to the
perturbative equations of motion (\ref{eq:explicit gauge
invariants})--(\ref{eq:brane wave equation}).  Second, there is the
effect of the bulk degrees of freedom encapsulated by the bulk
master variable $\Omega$ (or, equivalently, the KK fluid variables
$\Delta_\KK$, $V_\KK$ and $\delta\pi_\KK$).  To separate out the two
effects, it is useful to introduce the 4-dimensional effective
theory discussed above where all $\mathcal{O}(\rho/\sigma)$
corrections to GR are retained, but the bulk effects are removed by
artificially setting $\Omega = 0$. In the case of radiation
domination, we obtain equations for the effective theory density
contrast $\Delta_\eff$ and curvature perturbation $\zeta_\eff$ from
(\ref{eq:explicit zeta}) and (\ref{eq:brane wave equation}) with
$\Omega_\b = 0$:
\begin{subequations}
\begin{align}
    0 & = \frac{d^2\Delta_\eff}{d\eta^2} + \left( \frac{k^2}{3} -
    \frac{4\rho a^2}{\sigma\ell^2} - \frac{18\rho^2
    a^2}{\sigma^2\ell^2}\right) \Delta_\eff, \\
    \zeta_\eff & = \left( \frac{1}{4}
    + \frac{3 \rho a^2}{2 \sigma k^2 \ell^2} + \frac{9\rho^2
    a^2}{4\sigma^2k^2\ell^2}\right)\Delta_\eff  + \frac{3Ha}
    {4k}\frac{d\Delta_\eff}{d\eta}.
\end{align}
\end{subequations}
These in conjunction with the Friedmann equation
(\ref{eq:Friedmann}) and the conservation of stress energy
(\ref{eq:conservation}) give a closed set of ODEs on the brane
that describe all of the $\mathcal{O}(\rho/\sigma)$ corrections to
GR.

\begin{figure}
\begin{center}
    \includegraphics{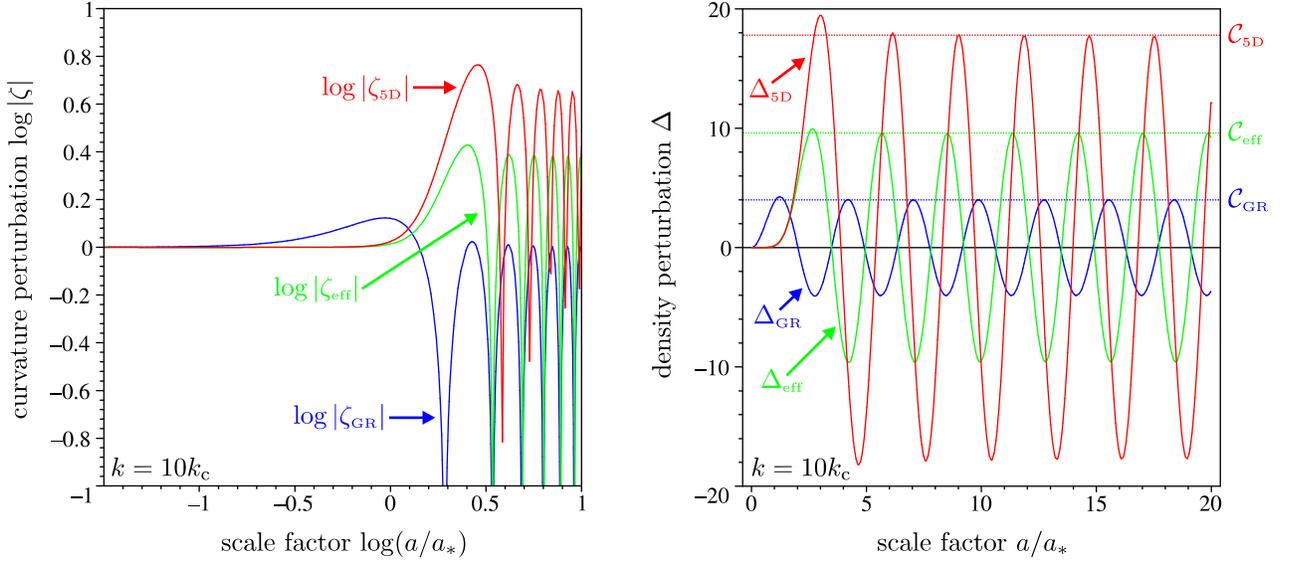} \caption{A comparison of the
    behaviour of the curvature perturbation $\zeta$ (\emph{left}) and the density
    perturbation $\Delta$ (\emph{right}) in the full 5-dimensional theory including
    KK contributions (5D), the effective 4-dimension theory including $\mathcal{O}(\rho/\sigma)$
    corrections (eff), and ordinary general relativity (GR).  The waveforms for each theory are normalized
    such that $\zeta = 1$ on superhorizon scales.  We can clearly see that given the
    same primordial values of the curvature perturbation, the final amplitude of the
    density perturbation in the 5D theory $\mathcal{C}_\fiveD$ is larger than in the
    effective theory $\mathcal{C}_\eff$, which in turn is larger than the GR
    value $\mathcal{C}_\GR = 4$.\label{fig:effective}}
\end{center}
\end{figure}

In Fig.~\ref{fig:effective}, we plot the predictions of GR, the
effective theory, and the full 5-dimensional simulations for the
behaviour of $\zeta$ and $\Delta$ for a supercritical mode.  Since
in any given model we expect the primordial value of the curvature
perturbation to be fixed by inflation, it makes physical sense to
normalize the waveforms from each theory such that
\begin{equation}\label{eq:zeta normalization}
    \zeta_\fiveD \approx \zeta_\eff \approx \zeta_\GR \approx 1,
    \quad a \ll a_*.
\end{equation}
When this is enforced we see that the effective theory predicts a
larger final amplitude for the density perturbation than GR.
Furthermore, the final amplitude in the 5-dimensional simulation
is larger than both of the other theories.  From this we can infer
that both $\mathcal{O}(\rho/\sigma)$ and KK effects induce
enhancement in the amplitude of perturbations.  This is in
contrast to the situation for tensor perturbations in the
Randall-Sundrum model, where modification of the expansion serves
to increase the amplitude of gravitational waves while the bulk
effects tend to decrease it
\cite{Hiramatsu:2003iz,Hiramatsu:2004aa,Kobayashi:2005dd,Kobayashi:2005jx,%
Kobayashi:2006pe,Ichiki:2003hf,Seahra:2006tm}.

\begin{figure}
\begin{center}
    \includegraphics{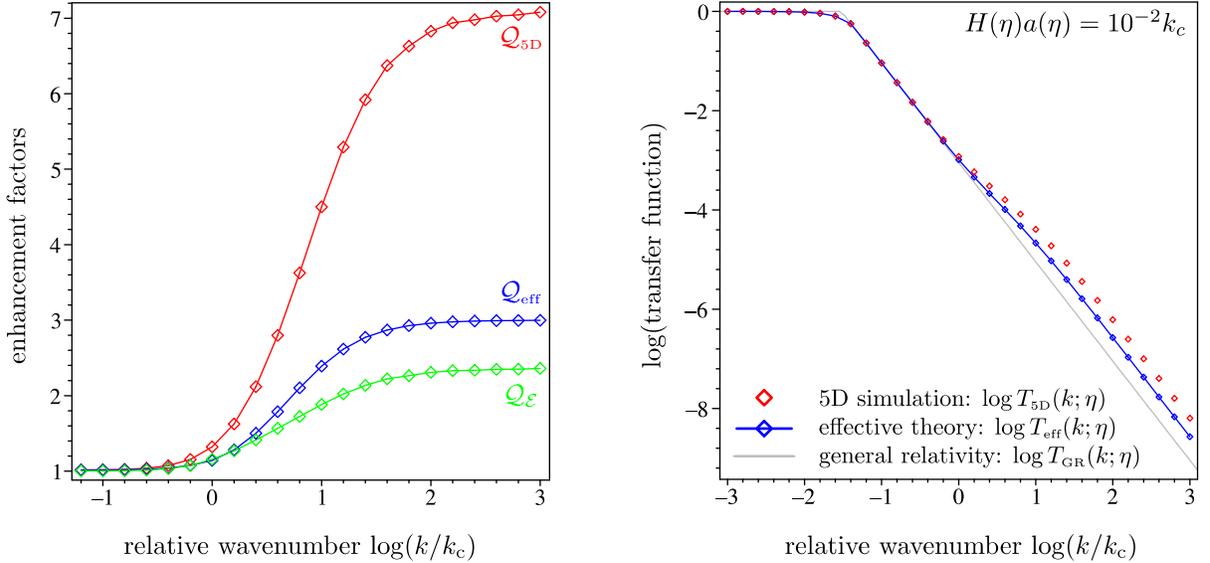} \caption{Density perturbation
    enhancement factors (\emph{left}) and transfer functions
    (\emph{right}) from simulations, effective theory, and general
    relativity.  All of the $\mathcal{Q}$ factors monotonically
    increase with $k/k_\c$, and we see that the $\Delta$ amplitude
    enhancement due to $\mathcal{O}(\rho/\sigma)$ effects $\mathcal{Q}_\eff$ is
    generally larger than the enhancement due to KK effects
    $\mathcal{Q}_\KK$.  For asymptotically small scales $k \gg k_\c$,
    the enhancement seems to level off.  The transfer functions in
    the right panel are evaluated at a given subcritical epoch in the
    radiation dominated era. The $T$ functions show how, for a fixed
    primordial spectrum of curvature perturbations
    $\mathcal{P}^\prim_\zeta$, the effective theory predicts excess
    power in the $\Delta$ spectrum $\mathcal{P}_\Delta \propto T^2
    \mathcal{P}^\prim_\zeta$ on supercritical/subhorizon scales
    compared to the GR result.  The excess small-scale power is even
    greater when KK modes are taken into account, as shown by
    $T_\fiveD(k;\eta)$. \label{fig:spectra}}
\end{center}
\end{figure}

The amount of enhancement should be a function of the frequency of
the mode, since we expect extremely subcritical modes $k \ll k_\c$
to behave as in GR.  To test this, we can define a set of
``enhancement factors'', which are functions of $k$ that describe
the relative amplitudes of $\Delta$ after horizon crossing in the
various theories.  As in Fig.~\ref{fig:effective}, let the final
amplitudes of the density perturbation with wavenumber $k$ be
$\mathcal{C}_\fiveD(k)$, $\mathcal{C}_\eff(k)$ and
$\mathcal{C}_\GR(k)$ for the 5-dimensional, effective and GR
theories, respectively, given that the normalization (\ref{eq:zeta
normalization}) holds. Then, we define enhancement factors as
\begin{equation}
    \mathcal{Q}_\eff(k) =
    \frac{\mathcal{C}_\eff(k)}{\mathcal{C}_\GR(k)}, \quad \mathcal{Q}_\KK(k) =
    \frac{\mathcal{C}_\fiveD(k)}{\mathcal{C}_\eff(k)}, \quad \mathcal{Q}_\fiveD(k) =
    \frac{\mathcal{C}_\fiveD(k)}{\mathcal{C}_\GR(k)}.
\end{equation}
It follows that $\mathcal{Q}_\eff(k)$ represents the
$\mathcal{O}(\rho/\sigma)$ enhancement to the density perturbation,
$\mathcal{Q}_\KK(k)$ gives the magnification due to KK modes, while
$\mathcal{Q}_\fiveD(k)$ gives the total 5-dimensional amplification
over the GR case. These enhancement factors are shown in the left
panel of Fig.~\ref{fig:spectra}.  We can see that they all increase
as the scale is decreased, and that they all approach unity for $k
\rightarrow 0$.  Since $\mathcal{Q} = 1$ implies no enhancement of
the density perturbations over the standard result, this means we
recover general relativity on large scales.  For all wavenumbers we
see $\mathcal{Q}_\eff
> \mathcal{Q}_\KK > 1$, which implies that the amplitude magnification due
to the $\mathcal{O}(\rho/\sigma)$ corrections is always larger than
that due to the KK modes.  Interestingly, the $\mathcal{Q}$-factors
appear to approach asymptotically constant values for large $k$:
\begin{equation}
    \mathcal{Q}_\eff(k) \approx 3.0,
    \quad \mathcal{Q}_\KK(k) \approx
    2.4,
    \quad \mathcal{Q}_\fiveD(k) \approx
    7.1, \quad k \gg k_\c.
\end{equation}
It is difficult to know if these are the true asymptotic limits for
$k \rightarrow \infty$ due to the limitations of computing power.

In cosmological perturbation theory, transfer functions are very
important quantities. They allow one to transform the primordial
spectrum of some quantity set during inflation into the spectrum
of another quantity at a later time.  In this sense, they are
essentially the Fourier transform of the retarded Green's function
for cosmological perturbations.  There are many different transfer
functions one can define, but for our case it is useful to
consider a function $T(k)$ that will tell us how the initial
spectrum of curvature perturbations $\mathcal{P}_\zeta^\prim$ maps
onto the spectrum of density perturbations $\mathcal{P}_\Delta$ at
some low energy epoch within the radiation era characterized by
the conformal time $\eta > \eta_\c$. It is customary to normalize
transfer functions such that $T(k;\eta) \approaches{k}{0} 1$,
which leads us to the following definition
\begin{equation}
    T(k;\eta) = \frac{9}{4} \left[ \frac{k}{H(\eta)a(\eta)}
    \right]^{-2} \frac{\Delta_k(\eta)}{\zeta^\prim_k}.
\end{equation}
Here, $\zeta^\prim_k$ is the primordial value of the curvature
perturbation and $\Delta_k(\eta)$ is the maximum amplitude of the
density perturbation in the epoch of interest.  As demonstrated in
Fig.~\ref{fig:subcritical}, we know that we recover the GR result
in the extreme small scale limit
\begin{equation}
    \Delta_k(\eta) \approaches{k}{0} \frac{4}{9} \left[ \frac{k}{H(\eta)a(\eta)}
    \right]^{2} \zeta^\prim_k, \quad (a > a_c),
\end{equation}
which gives the transfer function the correct normalization.  In
GR, the transfer function is accurately given by
\begin{equation}\label{eq:GR transfer function}
    T_\GR(k;\eta) \approx
    \begin{cases}
        1, & k < 3Ha, \\ (3Ha/k)^2, & k > 3Ha.
    \end{cases}
\end{equation}
In the righthand panel of Fig.~\ref{fig:spectra}, we show the
transfer functions derived from GR, the effective theory and the
5-dimensional simulations.  As expected, the $T(k;\eta)$ for each
formulation match one another on subcritical scales $k < k_\c$.
However, on supercritical scales we have $T_\fiveD > T_\eff >
T_\GR$.

Note that if we are interested in the transfer function at some
arbitrary epoch in the low-energy radiation regime $Ha \gg k_c$, it
is approximately given in terms of the enhancement factor as
follows:
\begin{equation}\label{eq:5D transfer function}
    T_\fiveD(k;\eta) \approx
    \begin{cases}
        1, & k < 3Ha, \\ (3Ha/k)^2 \mathcal{Q}_\fiveD(k), & k > 3Ha,
    \end{cases}
\end{equation}
Now, the spectrum of density fluctuations at any point in the
radiation era is given by
\begin{equation}
    \mathcal{P}_\Delta(k;\eta) = \frac{16}{81} T^2(k;\eta)
    \left( \frac{k}{Ha} \right)^4 \mathcal{P}^\prim_\zeta(k).
\end{equation}
Using the enhancement factor results in Fig.~\ref{fig:spectra} with
Eqns.~(\ref{eq:GR transfer function}) and (\ref{eq:5D transfer
function}), we see that the RS matter power spectrum (evaluated in
the low-energy regime) is $\sim 50$ times bigger than the GR
prediction on supercritical scales $k \sim 10^3 k_\c$.

\subsection{Alternate initial data}\label{sec:initial data}

Before concluding, we wish to briefly revisit the issue of initial
conditions. We have argued above that the existence of a dominant
growing mode at high-energy and on superhorizon scales means that
the details of initial data for our simulations are unimportant. To
some degree, the fact that the PS and CI algorithms produce
essentially identical results is a good confirmation of this, since
each method uses a different prescription for initial data.

\begin{figure}
\begin{center}
    \includegraphics{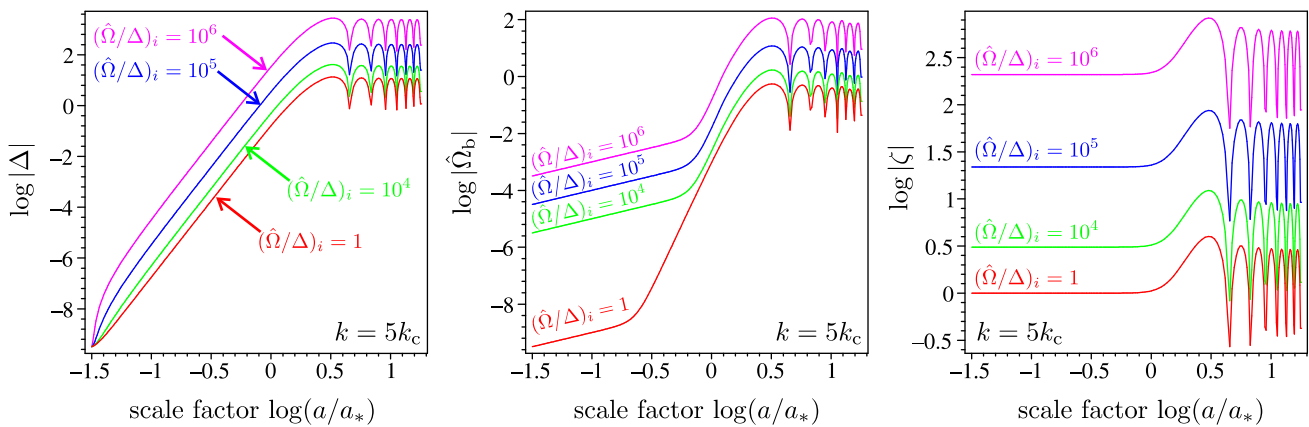} \caption{Simulated waveforms for
    $\hat\Omega =$ constant initial data in the CI method.  Here,
    $(\hat\Omega/\Delta)_i$ is the ratio of the bulk field to the density
    perturbation at the initial time; this ratio is zero for the other simulations
    in this paper.  The early time behaviour of $\Delta$ and $\hat\Omega_\b$ shows some
    sensitivity to $(\hat\Omega/\Delta)_i$, but otherwise the waveforms are
    identical to one another up to an overall amplitude scaling.  Indeed, if we normalize
    the simulation results by the initial value of $\zeta$ we find that
    all of the $\Delta$ waveforms are coincident at late times, which means that
    the enhancement factor $\mathcal{Q}_\fiveD$ and transfer function
    $T_\fiveD$ are insensitive to $(\hat\Omega/\Delta)_i$.  Note that
    at early times we have $\hat\Omega_\b \propto a$, which matches
    the behaviour of the subdominant growing mode from Eq.~(\ref{eq:high-energy solutions}).
    This confirms our expectations: varying the initial data causes the other subdominant modes
    to be excited to various degrees, but the dominant growing mode always ``wins''
    at late times.\label{fig:init data}}
\end{center}
\end{figure}

But what if we were to use different classes of initial data? For
example, we could modify the CI initial data such that the bulk
field has constant amplitude along the initial data surface:
\begin{equation}
    \hat\Omega(\di\mathcal{M}_\CI^-) = \hat\Omega_i =
    \text{constant}.
\end{equation}
The waveforms generated by such a choice are shown in
Fig.~\ref{fig:init data}.  Though the overall amplitude and early
time behaviour of the signal seems to be sensitive to the initial
value of $\hat\Omega$, the ratio of the final amplitude of $\Delta$
to the initial value of $\zeta$ is the same for each choice of
initial data.  We have confirmed that this is also true for other choices
of $\hat\Omega(\di\mathcal{M}_\CI^-)$, including the case when the
initial data is a simple sinusoid.\footnote{Sinusoidal initial data
was used in Ref.~\cite{Seahra:2006tm} to test the insensitivity of
the final spectrum of the stochastic background of gravitational
waves to initial conditions in RS models.} Ultimately, it is the
ratio $\Delta(k\eta \gg 1)/\zeta(k\eta \ll 1)$ that is relevant to
the transfer function used to transcribe the predictions of
inflation into observable quantities, hence we can be confident that
our principal results hold for reasonable modifications of initial
data.

\section{Discussion and implications}\label{sec:discussion}

In this paper, we have written down the equations of motion for
generic gauge-invariant scalar cosmological perturbations in the RS
braneworld model in a form suitable for numeric analysis
(\S\ref{sec:scalar perturbations}).  We have developed new analytic
approximations for the behaviour of fluctuations on high-energy
superhorizon scales (\S\ref{sec:asymptotic}) for a radiation
dominated brane.  We have applied two different numerical algorithms
to solve the equations in the early radiation-dominated universe
(\S\ref{sec:numeric}). Our numerical results show that the amplitude
of modes which enter the Hubble horizon during the high-energy
regime gets enhanced over the the standard GR result. Conversely,
modes which enter at low energy do not show any late-time deviations
from standard theory, as seen in Fig.~\ref{fig:subcritical}.  We
only recover standard results for the metric perturbations, such as
$\Psi \approx -\Phi$, at low energies.

Our simulations confirm the common sense expectation that all
tangible effects from the fifth dimension are on scales smaller than
a critical value $k_c^{-1}$, whose physical size today is given by
Eq.~(\ref{eq:scale today}). To parameterize the degree of
amplification as a function of scale, we introduced so-called
``enhancement factors'' and transfer functions in
\S\ref{sec:transfer}.  These show that the degree of enhancement
over the GR case seems to reach a constant value at high $k$; the
amplification of the subhorizon matter power spectrum is $\sim 50$
for $k \sim 10^3 k_c$.  We presented analytic arguments and
numerical evidence that our results are robust against modifications
of the initial data for simulations (\S\ref{sec:initial data}).

As discussed in \S\ref{sec:introduction} and \S\ref{sec:transfer},
the enhancement of perturbations upon horizon crossing can be
attributed to two effects: namely, the $\mathcal{O}(\rho/\sigma)$
corrections to the background dynamics and the influence of the KK
modes.  Both of these give roughly equal contributions to the
enhancement (Fig.~\ref{fig:spectra}).  One could have anticipated
the KK enhancement on simple physical grounds by arguing as follows:
In Ref.~\cite{Garriga:1999yh} it was shown that the gravitational
force of attraction between two bodies in the RS one brane model is
\begin{equation}
    \text{gravitational force} \sim \begin{cases}
    {\kappa_4^2}/{r^2}, & r \gg \ell, \\
    {\kappa_5^2}/{r^3} = {\kappa_4^2 \ell}/{r^3}, &
    r \ll \ell.
    \end{cases}
\end{equation}
%The $1/r^4$ correction term is the direct effect of the KK degrees
%of freedom on Newton's law.
%KK
That is, the Newtonian force becomes 5-dimensional on small scales.
%KK
This implies that the gravitational
force is stronger than usual on scales $r \lesssim \ell$. It
then follows that the self-gravity of perturbative modes that
enter the horizon at high energies $H\ell \gtrsim 1$ will be
greater than those which enter at low energies $H\ell \lesssim 1$.
Therefore, we should expect that the amplitude of small-scale
modes to be magnified over the amplitude of large-scale modes
during horizon crossing in braneworld cosmology, which is exactly
what we have seen in our simulations.\footnote{However, we should note
that this simple argument neglects the influence of the KK
anisotropic stress, which we know is non-negligible at high
energies.}

The amplitude enhancement of perturbations is important on comoving
scales $\lesssim 10 \,\text{AU}$, which are far too small to be
relevant to present-day/cosmic microwave background measurements of
the matter power spectrum. However, it may have an important bearing
on the formation of compact objects such as primordial black holes
and boson stars at very high energies, i.e. the greater
gravitational force of attraction in the early universe will create
more of these objects than in GR (different aspects of primordial
black holes in RS cosmology in the context of various effective
theories have been considered in
Refs.~\cite{Guedens:2002km,Guedens:2002sd,Majumdar:2002mr,Clancy:2003zd,%
Sendouda:2003dc,Sendouda:2004hz,Sendouda:2006nu}). We know that the
abundance of primordial black holes can be constrained by big bang
nucleosythesis and observations of high-energy cosmic rays, so it
would be interesting to see if the kind of enhancement of density
perturbations predicted in this paper can be used to derive new
limits on Randall-Sundrum cosmology.  We leave this issue for future
work.

\begin{acknowledgments}
AC is supported by FCT (Portugal) PhD fellowship SFRH/BD/19853/2004.
KK and SSS are supported by PPARC (UK).
\end{acknowledgments}

\bibliography{ms}

\end{document}